\documentstyle[twoside,fleqn,espcrc2]{article} 

\newcommand{\AmS}{{\protect\the\textfont2
  A\kern-.1667em\lower.5ex\hbox{M}\kern-.125emS}}

\title{Phenomenology of neutrino physics in the Kaluza-Klein theories of
low scale gravity}

\author{A. N. Ioannisian\address{Instituto de Fisica Corpuscular -        
CSIC - Univ. de Valencia, \\
Edificio Institutos de Paterna -
Apartado de Correos 2085, 
46071 Valencia, Spain }%
        \thanks{On leave from Yerevan Physics Institute, Alikhanyan Br. 2,
 Yerevan 375036, Armenia }}       
\begin{document}

\begin{abstract}
We discuss the phenomenological consequences of theories which describe
sterile neutrinos in large extra dimensions.
We show that the Kaluza-Klein tower of
the singlet neutrinos, albeit tiny individual contribution in electroweak
processes, act cumulatively, giving rise to non-universality of the weak
interactions of the light neutrinos and to flavour-violating radiative
processes.  Owing to these non-decoupling effects of th Kaluza--Klein
neutrinos, we derive strong constraints on the parameters of the theory
that originates from the non-observation of flavour-violating and
universality-breaking phenomena. In this theory we propose a four-neutrino
model which can reconcile the existing data coming from underground
experiments in terms of neutrino oscillations, together with the hint from
the LSND experiment and a possible neutrino contribution to the hot dark
matter of the Universe. 
\end{abstract}

\maketitle


Recently, it has been proposed \cite{ADD} the radical possibility that the
observed small value of the gravitational constat at long distance is
ascribed to the spreading of the gravitational force in $\delta$ large
extra spatial dimensions. It follows that the Standard Model (SM) fields
are confined to a 3-brane configuration, while the large compactified
dimensions are probed only by gravity and singlet, under SM gauge group,
fields.

In this  presentation we examine  the phenomenological
consequences of higher - dimensional  isosinglet neutrinos on  collider
and lower energy experiments.

For our phenomenological study we adopt a variant \cite{AP} of the
model discussed in Ref.\ \cite{ADDM} \cite{DDG}.  For definiteness, we
consider a model that minimally extends the SM-field content by
one singlet Dirac neutrino, $N(x,y)$, which propagates in a
$[1+(3+\delta)]$-dimensional Minkowski space. The $y$-coordinates are
compactified on a circle of radius $R$ by applying the periodic
identification: $y \equiv y + 2\pi R$.

The relevant part of the Lagrangian of our minimal model reads
\begin{eqnarray}
  \label{Leff}
&&\int\limits_0^{2\pi R}\!\! d\vec{y}\
 \Big[\, \bar{N} \Big( i\gamma^\mu \partial_\mu\, +\,
 i\gamma_{\vec{y}} \partial_{\vec{y}} -m \Big) N\  \nonumber \\
&&+\ \delta (\vec{y}) \ \Big( \sum_{l=e,\mu,\tau}\,
\frac{h_l}{M_F^{\delta/2}} L_l\tilde{\Phi}
\xi\, +\, {\rm H.c.}\,\Big) \,\Big],
\end{eqnarray}
where $\tilde{\Phi}$ ( $<\tilde{\Phi}> =v=174$Gev ) and $L_l$ are higgs
and lepton doublets, respectively.
We have assumed that only one two-component spinor
$\xi$ from the high dimensional spinor $N$ couples to our 3 brane.

We can now express the two-component
spinor $\xi$ and its dirac partner $\eta$ from high dimensional
spinor $N$ in terms of a Fourier series expansion as follows:
\begin{eqnarray}
  \label{xi}
&&\left[ \xi (x,y), \ \eta (x,y)\right]  =  \\
&&\frac{1}{(2\pi
R)^{\delta/2}}\
\sum_{\vec{n}}\,
\left[ \xi_{\vec{n}}(x), \ \eta_{\vec{n}}(x) \right] \
\exp\bigg(\frac{i\vec{n}\vec{y}}{R}\bigg).
\nonumber
\end{eqnarray}
Substituting into the Eq.  (\ref{Leff})  and then performing the $\vec{y}$
integration
yields
\begin{eqnarray}
  \label{LKK}
&&\sum_{\vec{n}}
\bigg\{ \bar{\xi}_{\vec{n}}
( i\bar{\sigma}^\mu \partial_\mu) \xi_{\vec{n}} +
\bar{\eta}_{\vec{n}} ( i\bar{\sigma}^\mu \partial_\mu) \eta_{\vec{n}}
- \\
&&\Big[ \xi_{\vec{n}} \Big( (m + \frac{i|\vec{n}|}{R}) \eta_{-{\vec{n}}}
+\sum_{l=e,\mu,\tau}
\bar{h}_l L_l\tilde{\Phi} \Big) + {\rm H.c.} \Big] \bigg\}
\nonumber
\end{eqnarray}
where
$\bar{h}_l = h_l \frac{M_F}{M_{\rm P}}$ , $|\vec{n}|^2\equiv
\sum_{i=1}^{\delta} n_i^2$.

As it is shown in Eq.  (\ref{LKK}), there are small mixings mass
terms, $\bar{h_l}v$, between any KK state $\eta_{\vec{n}}$ and left
neutrinos. It was firstly noticed in \cite{ADDM,DDG}, the four-dimensional
Yukawa couplings $\bar{h}_l$ are naturally suppressed by the volume factor
$M_F/M_{\rm P}$.

After diagonalizing the mass matrix
these terms give rise to the small admixture of the left neutrinos
to the heavy KK states, $\eta_{\vec{n}}^\prime$, with the mass
$\sqrt{m^2+\vec{n}^2/R^2}$
\begin{equation}
\eta_{\vec{n}}^\prime \approx \eta_{\vec{n}} +B_{l,\vec{n}}\nu_L^l, \
 B_{l,\vec{n}} \approx {h}_l
\frac{v M_F}{M_P\sqrt{m^2+\vec{n}^2/R^2}}.
\end{equation}

Applaying to the unitarity condition we arrive to the following light
states

\begin{equation}
\label{nonuni}
\nu^l_{light} \approx \frac{1}{\sqrt{1+\sum_{\vec{n}}|B_{l,\vec{n}}|^2}}
( \nu_L^l + \sum_{\vec{n}} B_{l,\vec{n}}\eta_{\vec{n}}^\prime ).
\end{equation}

The most striking feature  of the  higher-dimensional scenario, as it
follows from  Eq.(\ref{nonuni}), is  the
loss of the lepton universality  in  the electroweak processes.

It is useful  to define the
mixing parameters
\begin{eqnarray}
&&  \label{snul} (s^{\nu_l}_L)^2 \equiv \sum_{\vec{n}} |B_{l,\vec{n}}|^2
\approx h^2_l\frac{v^2}{M_F^2}\sum_{\vec{n}} \frac{M_F^4
M_P^{-2}}{(m^2+\frac{\vec{n}^2}{R^2})} 
\nonumber \\
&&\approx \left\{ \begin{array}{l}
\frac{\displaystyle \pi h^2_l v^2}{\displaystyle M^2_F}\,
\ln\bigg(\displaystyle \frac{M^2_F}{m^2}+1\bigg), 
\delta=2 \\ \\ \frac{\displaystyle S_\delta}{\displaystyle \delta - 2}\,
\frac{\displaystyle h^2_l v^2}{\displaystyle M^2_F}, 
\delta > 2\, .  \end{array}\ \right.
\end{eqnarray}
In order to evaluate the sumation in the above equation we approximate the
sum over the KK states by an integral, which has an upper ultra-violet
(UV)  cutoff at $M_F R$, above which
string-threshold effects are expected to become more important.

As  can be seen from  Table \ref{Tab1},  the mixings $(s^{\nu_l}_L)^2$
may be constrained by a  number of new-physics observables induced  at
the tree  level.  These observables measure  possible non-universality
effects in $\mu$, $\tau$ and $\pi$ decays.   In this respect, in Table
\ref{Tab1} we have defined  $R_\pi =\Gamma(\pi \to e\nu)/\Gamma(\pi\to
\mu\nu)$  and  $R_{\tau \mu}=B(\tau  \to e   \nu \nu)/B(\tau  \to  \mu
\nu\nu)$.

\begin{table*}[t]
\caption{Tree level limits on $M_F/$TeV from
non-universal couplings of the neutrinos to $W$ and $Z$ bosons.}
\label{Tab1}
\begin{center}
\begin{tabular}{c|cc|cc}
\hline
 & $h_e\ =\ h_\mu$ &\hspace{-1cm}$=\ h_\tau\ =\ h$ &
\hspace{1.cm}$h_\mu =0$ &\hspace{-0.5cm}and $h_e = h_\tau$ \\
Observable & Lower limit &
\hspace{-0.6cm} on $M_F/h$ & Lower limit
&\hspace{-0.6cm} on $M_F/h_\tau$ \\
& $\delta =2$ & $\delta > 2$ & $\delta =2 $ & $\delta > 2$ \\
\hline
$1 - \frac{\displaystyle \Gamma(\mu\to e\nu\nu )}{\displaystyle
\Gamma_{\rm SM} (\mu\to e\nu\nu )}$
 & $8.9\,\ln^{1/2}\frac{\displaystyle M_F}{\displaystyle m}$ &
$\frac{\displaystyle 3.5\, S^{1/2}_\delta}{\displaystyle \sqrt{\delta -2}
}$&
$6.3\,\ln^{1/2}\frac{\displaystyle M_F}{\displaystyle m}$&
$\frac{\displaystyle 2.5\, S^{1/2}_\delta}{\displaystyle \sqrt{\delta -2}
}$\\
\hline
$1 - \frac{\displaystyle \Gamma(Z\to \nu\nu )}{\displaystyle
\Gamma_{\rm SM} ( Z\to \nu\nu )}$
 & $5.9\,\ln^{1/2}\frac{\displaystyle M_F}{\displaystyle m}$ &
$\frac{\displaystyle 2.4\, S^{1/2}_\delta}{\displaystyle \sqrt{\delta
-2}}$&
$4.8\,\ln^{1/2}\frac{\displaystyle M_F}{\displaystyle m}$&
$\frac{\displaystyle 1.9\, S^{1/2}_\delta}{\displaystyle \sqrt{\delta
-2}}$\\
\hline
$1 - \frac{\displaystyle R_\pi }{\displaystyle R^{\rm SM}_\pi }$
        & $-$ & $-$ &
$18.7\,\ln^{1/2}\frac{\displaystyle M_F}{\displaystyle m}$&
$\frac{\displaystyle 7.5\, S^{1/2}_\delta}{\displaystyle \sqrt{\delta
-2}}$\\
\hline
$1 - \frac{\displaystyle R_{\tau \mu}}{\displaystyle R^{\rm SM}_{\tau \mu}
}$
 & $-$ & $-$ & $5.7\,\ln^{1/2}\frac{\displaystyle M_F}{\displaystyle m}$&
$\frac{\displaystyle 2.3\, S^{1/2}_\delta}{\displaystyle \sqrt{\delta
-2}}$\\
\hline
\end{tabular}
\end{center}
\end{table*}
\begin{table*}[t]
\caption{One-loop-level limits on $M_F/h^2$.}
\label{Tab2}
\begin{tabular}{cccc}
\hline
 & $h_e\ =\ h_\mu$ &\hspace{-1cm}$=\ h_\tau\ =h \ge\ 1$ & \\
Observable & Lower limit &
\hspace{-0.1cm} on $M_F/h^2 \ [\,{\rm TeV}\,]$& \\
& $\delta =2$ & $\delta = 3$ & $\delta =6 $  \\
\hline
Br$(\mu \to e \gamma)$ & $75$ & $43$ & $33$\\
\hline
Br$(\mu \to e e e)   $ & $250$ & $230$ & $200$\\
\hline
Br$(\mu \ ^{48}_{22}{\rm Ti} \to e \ ^{48}_{22}{\rm Ti})$ & $380$ & $320$
&300\\
\hline
\end{tabular}
\end{table*}

Owing to the tower of the KK singlet neutrinos which acts cumulatively
in   the loops,     significant  universality-breaking  as   well   as
flavour-violating  effects   are  induced  in   electroweak  processes
involving  $\gamma$-\cite{FP} and $Z$-  boson~\cite{IN}  interactions.
In particular, as has been explicitly shown recently,\cite{IN} we find
that the  cumulative presence of the KK  states leads to  an effective
theory whose  Yukawa interactions are  mediated  by order-unity Yukawa
couplings of the  original Lagrangian before compactification. In this
case,  we    expect  a higher-dimensional    non-decoupling phenomenon
analogous to the  one studied earlier  in renormalizable 4-dimensional
theories.\cite{IP,PS}      For      example,       the       effective
lepton-flavour-violating vertex $Zll'$   that occurs in $\mu \to  eee$
exhibits  the  dependence:   ${\cal   T}(Zl'l)  \propto h_l   h_{l'}\,
(v^2/M^2_F)\,  \sum_{k=e,\mu,\tau}\,  (h^2_k   v^2)/M^2_W$, i.e.\  its
strength  increases  with the fourth   power of the higher-dimensional
Yukawa  couplings.  This  should   be contrasted  with the  respective
photonic  amplitude   ${\cal  T}(\gamma   l'l) \propto    h_l h_{l'}\,
v^2/M^2_F$, whose strength increases only quadratically.

Based on this cumulative non-decoupling  effect, we are able to derive
strong limits  on  the $M_F/h^2$, for $h\ge   1$.  As is  displayed in
Table~\ref{Tab2}, the   strongest  limits  are   obtained from    $\mu
\not\!\to eee$ and  the absence of  $\mu$-to-$e$ conversion in nuclei.
Of  course,  the    limits  presented  here  contain    some degree of
uncertainty, which is  inherent  in all effective   non-renormalizable
theories with a   cut-off  scale, such  as  $M_F$.   Nevertheless, our
results  are very  useful, since   they indicate the  generic  size of
constraints that one has to encounter in model-building considerations
with sterile neutrinos.\cite{KKneutr,IV}

Now we concentrate in the minimum brane-inspired scheme in which all
elements required to explain the neutrino anomalies (the LSND/HDM as well
as the solar and atmospheric mass scales) are generated by the physics of
extra dimensions \cite{IV}. For definiteness, we will be considering a
model that minimally extends the standard field content by one bulk
neutrino, $N(x,y)$ (with zero high dimensional mass term), singlet under
the SM gauge group. This propagates on a $[1+(3+n)]$-dimensional Minkowski
space with $\delta \ge n$.

After
integrating out
heavy Kaluza-Klein states,
the effective Lagrangian has the following form 
\begin{equation}
  \label{LLKK}
{\cal L}_{\rm eff}  =  {\cal L}_{\rm SM} \
+\, \Big(\, \sum_{l=e,\mu,\tau}\,\bar{h}_l\, L_l\tilde{\Phi} \xi_0\  +\
{\rm H.c.}\,\Big)\, ,
\end{equation}
where $\xi_0$ is the zero mode of the Kaluza-Klein states and now
\begin{equation}
  \label{hln}
\bar{h_l}\ =\  \bigg(\frac{M_F}{M_{\rm P}}\bigg)^{\frac{n}{\delta}}\
h_l .
\end{equation}
In order to account for the LSND or hot dark matter mass scale $m_\nu
\sim 1$ eV or so, we choose $\delta=6$ and $n=4$, giving $\bar{h}_l \sim
10^{-10} h_l$.

Now we turn to the Majorana masses for neutrinos. These are crucial in
order to generate the mass splittings required in neutrino oscillation
interpretations of the solar and atmospheric neutrino anomalies found
in underground experiments. As shown in Ref.~\cite{ADDM} the
neutrinos may get small Majorana masses via interactions with distant
branes where fermion number is maximally broken. 

 The Majorana part of the neutrino
masses is then expected to be
\begin{equation}
\label{majorana}
m_{l l^\prime} \sim f_{l l^\prime}\frac{v^2}{M_F}
\left(\frac{M_F}{M_{\rm
P}}\right)^{2n/\delta - 4/\delta}
\label{light}
\end{equation}

The neutrino mass matrix takes in the basis ($\nu_e, \nu_\mu, \nu_\tau,
\nu_s$ ) ($\nu_s=\xi_0$) the form

\begin{equation}
      \cal M_{\nu} ~=~ \pmatrix{
         m_{l l^\prime} &  M_l           \cr
         M^T_{l^\prime}     &  0     \cr}~.
\label{massmatrix}
\end{equation} 
Here $M_l = \bar{h}_l v$. Note that the $m_s$ entry in eq.
\ref{massmatrix} has been omitted since the bulk sector where the sterile
neutrino $\nu_s$ lives is eight--dimensional.

In the limit that Dirac mass terms ($M_l$) are much bigger than
Majorana mass terms ($m_{l l^\prime}$), two of the neutrinos are
massless and other two form Dirac state with a mass

\begin{equation}
\label{dirac}
m_\nu \equiv m_{LSND/HDM} = \sqrt{M^2_e+M^2_\mu+M^2_\tau}
\end{equation}

This state is identified by two angles $\theta$ and $\varphi$ defined as

\begin{equation}
\label{thetaphi}
\sin \theta = \frac{M_e}{m_\nu} , \ \ \tan \varphi = \frac{M_\mu}{M_\tau}.
\end{equation}

The entries $m_{l l^\prime}$ only arise due to the breaking lepton
number on distant branes. In the case $\delta = 6$ and $n=4$ they are
suppressed compared to the Dirac mass terms by the factor
$\frac{v}{M_F}\frac{f_{l l^\prime}}{h_l }$.  These terms give masses
to the lowest-lying neutrinos and also responsible for splitting Dirac
state to two Majorana states. For suitable values of the parameters,
these are in the right range to have a solution for solar and
atmospheric neutrino deficit. More especifically, from the latest fits
one needs \cite{Gonzalez-Garcia:2000sq}
\begin{equation}
\Delta m^2_{atm} \simeq 3.5 \times 10^{-3} eV^2
\end{equation}
in order to account for the full set of atmospheric neutrino data.

On the other hand the latest global analysis of solar neutrino data
characterized by
the best-fit points \cite{Gonzalez-Garcia:2000sq}
\begin{equation}
\Delta m^2_{LMA} \simeq 3.6 \times 10^{-5} eV^2
\end{equation}
\begin{equation}
\Delta m^2_{SMA} \simeq 5 \times 10^{-6} eV^2
\end{equation}
Now assuming naturalness, namely, that masses and splittings are of
the same order $\Delta m_{atm} \simeq m_{l l^\prime} \simeq
\sqrt{\Delta m^2_{\odot}}$ and since $\Delta m^2_{atm} \simeq 2
m_\nu \Delta m_{atm}$, one finds
\begin{equation}
m_\nu \simeq 0.8 eV {\rm for} \ {\rm LMA}
\end{equation}
\begin{equation}
m_\nu \simeq 0.3 eV {\rm for} \ {\rm SMA}
\end{equation}
characterizing the order of magnitude of the LSND/HDM scale in the LMA/SMA
cases.

Of course since clearly the solar mass splitting need not coincide
exactly with the lightest state masses, the above estimates are meant
to be crude order-of-magnitude estimates only. As a result the
LSND/HDM scales both in the LMA and in the SMA case can be larger than
estimated above.

If we assume that muonic neutrino coupling to the high dimensional
spinor is dominant ($h_e, h_\tau \ll h_\mu \simeq 0.1$ ) the light
sterile neutrino, $\nu_s$, combines with $\nu_\mu$ and form a
Quasi--Dirac state, crucial to account for the hint coming from the
LSND experiment, and may also contribute to the hot dark matter of the
Universe.
Apart from the mass of the Quasi--Dirac state~\cite{Valle:1983dk}, one
has the splittings between its components, as well as the masses of
two light active states.
The splittings between the heavy states and that characterizing the
lighter neutrinos will be associated with the explanations of the
atmospheric and solar neutrino anomalies, respectively.
The atmospheric neutrino deficit is ascribed to the $\nu_\mu$ to
$\nu_s$ oscilations.  The solar neutrino problem could be solved via
MSW small or large angle $\nu_e$ to $\nu_\tau$ solutions. This
reproduces exactly the phenomenological features of the model proposed
in ref.~\cite{PTV92}, providing a complete scenario for the present
neutrino anomalies.

\vspace{1cm}

\noindent{\bf Acknowledgements:} It is pleasure to thanks the
organizers for organizing a very interesting workshop. This work was
supported by Spanish DGICYT under grants PB98-0693 and SAB/1998-0136
(A.I.) and by the European Union TMR network ERBFMRXCT960090.

\end{document}